\documentclass[useAMS,usenatbib]{mn2e}
\usepackage{epsfig}
\usepackage{rotating}

\def\kms{km~s$^{-1}$}

\def\eg{{e.g.}}

\def\simgt{\hbox{\rlap{\raise 0.425ex\hbox{$>$}}\lower 0.65ex\hbox{$\sim$}}}
\def\simlt{\hbox{\rlap{\raise 0.425ex\hbox{$<$}}\lower 0.65ex\hbox{$\sim$}}}
\def\n891{NGC\,891}
\title[X-Ray Observations of NGC\,891 and SN1986J]
{X-Ray Observations of the edge-on star-forming galaxy NGC\,891 and its supernova SN1986J}
\author[R. F. Temple, S. Raychaudhury and I. R. Stevens]
{Rowan F. Temple\thanks{E-mail:
rft@star.sr.bham.ac.uk ; somak@star.sr.bham.ac.uk; irs@star.sr.bham.ac.uk }, 
Somak Raychaudhury and Ian R. Stevens\\
School of Physics and Astronomy, University of Birmingham, 
Edgbaston, Birmingham B15~2TT, England\\}

\begin{document}

\date{Accepted yyyy mm dd. Received yyyy mm dd; in original form yyyy mm dd}

\pagerange{\pageref{firstpage}--\pageref{lastpage}} \pubyear{2004}

\maketitle

\label{firstpage}

\begin{abstract}
We present {\it XMM-Newton} observations of NGC\,891, a nearby edge-on
spiral galaxy. We analyse the extent of the diffuse emission emitted
from the disk of the galaxy, and find that it has a single temperature
profile with best fitting temperature of 0.26~keV, though the
fit of a dual-temperature plasma with temperatures of 0.08 and 0.30~keV 
is also an acceptable fit.
There is a considerable amount of diffuse X-ray emission protruding from the
disk in the NW direction out to approximately 6~kpc.  We analyse the
point source population using a {\sl Chandra} observation, using a 
maximum likelihood method to find that the slope of the cumulative luminosity 
function of point sources in the galaxy is $-0.77^{+0.13}_{-0.1}$. 
Using a sample of other local galaxies, we compare the X-ray and
infrared properties of NGC\,891 with those of 'normal' and starburst
spiral galaxies, and conclude that NGC\,891 is most likely a starburst
galaxy in a quiescent state. We establish that the diffuse X-ray
luminosity of spirals scales with the far infra-red luminosity as
$L_{X}\propto L_{FIR} ^{0.87 \pm 0.07}$, except for extreme
starbursts, and NGC\,891 does not fall in the latter category. We
study the supernova SN1986J in both {\sl XMM-Newton} and {\sl Chandra}
observations, and find that the X-ray luminosity has been declining
with time more steeply than expected ($L_X\propto t^{-3}$).

\end{abstract}


\section{Introduction}

The environment of a spiral galaxy can influence its evolution in many
ways. Close interactions with nearby galaxies have been shown to
trigger inspire star formation, and thus alter the interstellar medium
(ISM) of the galaxy. Properties of the ISM could also be affected by
direct infall of material from the intergalactic medium (IGM), or, if
the galaxy is in a group or cluster, by the ram pressure stripping by
the IGM. However, in certain, outflows into the IGM could be a
significant mode of the interaction of a galaxy with its surroundings.
Bursts of star formation in the galaxy can, if sufficiently strong,
lead to the outflow of matter, which will be predominantly visible at
H$\alpha$ and X-ray wavelengths \citep{shopbell98,lehnert99}.  These
outflows can be strong enough to escape the visible extent of the
galaxy, but still be retained by the halo of the galaxy. In some of
these cases, the burst of star formation would be strong enough for
the outflow to be able to could escape the galaxy and its halo, and
appear as a superwind.

A detailed study of the region of interface between the optical extent
of the galaxy and the surrounding medium, is clearly
important. Edge-on spiral galaxies, such as the galaxy we study here,
NGC\,891, are oriented in such a way that these are particularly
useful in the study of this interface region \citep{b7}. Because of
the energies associated with supernovae and the wind of massive stars,
observations at X-ray energies are of particular interest in studying
the hot haloes or outflows from galaxies. Of the two currently
operating major X-ray missions, {\sl Chandra} has very high spatial
resolution and moderate collecting area, whereas {\sl XMM-Newton} has
reasonable spatial resolution, but a larger collecting area, and thus
should be able to see lower X-ray surface brightness features.

In this paper we will present results from {\sl XMM-Newton} and
Chandra observations of NGC\,891, an edge-on spiral galaxy. NGC\,891
is very similar in many respects to our own galaxy.  The high
inclination makes NGC\,891 ideal for studying the diffuse X-ray
emission from the outflow extending from the plane of the galaxy
(particularly in the NW direction), which has previously been studied
with the {\sl ROSAT} \citep{b55}, {\sl ASCA}
\citep{b5,b6} and {\sl Chandra} \citep{b7} observatories. 

High-resolution X-ray studies reveal diffuse emission up to
about $10^{37}$ erg~s$^{-1}$, but also reveal a rich point source
associated with stellar sources, such as
X-ray binaries and supernovae. The X-ray point source population 
is closely related with the history of star formation in the galaxy
\citep[\eg,][]{b13}.
The slope of the X-ray luminosity function (XLF) of point sources is
believed to be a good indicator of this, with flatter XLF slopes
indicating more recent star-formation (due to a large fraction of high
luminosity massive X-ray binaries). The point source populations
of several nearby galaxies have been analysed \citep{b15,b13,b37}. Of them,
\citet{b15} selected a small group of well-studied
spirals, and found that the XLF slope is generally steeper for normal
spiral galaxies than for the starburst galaxies, and correlated with
other observable parameters.  In this work, we compare these X-ray,
optical and IR properties of NGC\,891 with a similar sample of normal
and starburst galaxies.

As part of the point source population, the presence of an active
galactic nuclei (AGN) is also the subject of much debate in NGC\,891
and other galaxies, and how dominant or irrelevant the AGN is to the
energetics of the central regions of the galaxy. \citet{b7} show that
there is the possibility of a weak hard X-ray source (2.0-8.0~keV)
using the {\sl Chandra} data.  The discovery of a nuclear AGN in
NGC\,891 would help us to categorise it as a starburst galaxy.

A rare form X-ray point sources in galaxies are X-ray luminous young
supernovae.  There have been few X-ray detections of young supernovae,
and the ones that have been detected are associated with relatively
nearby Type II SN events (\cite{b56} and references within).  One of
these rare objects, detected at X-ray wavelengths, is SN1986J, located
in NGC\,891 and detected as a bright source in our X-ray
observations. It is also very radio bright. \citet{b56} studied the
evolution of SN1986J using both {\sl ASCA} and {\sl ROSAT}
data. SN1986J is still visible at X-ray energies and we present a new
analysis of the evolution of this source using both {\sl Chandra} and
{\sl XMM-Newton} data.

The paper is organised as follows: In \S2 we discuss the {\sl
XMM-Newton} and Chandra data used here, and present the reduction
procedure and results of the analysis of point sources (Chandra) and
the diffuse emission (XMM) in \S3.  In \S4, a comparison of X-ray and
near and far-IR properties is made with other nearby spiral
galaxies. Observations of the supernova SN1986J are presented in \S5,
and compared with observations at previous epochs. General conclusions
are presented in the final section.

\begin{figure*}
\begin{flushleft}
\epsfig{file=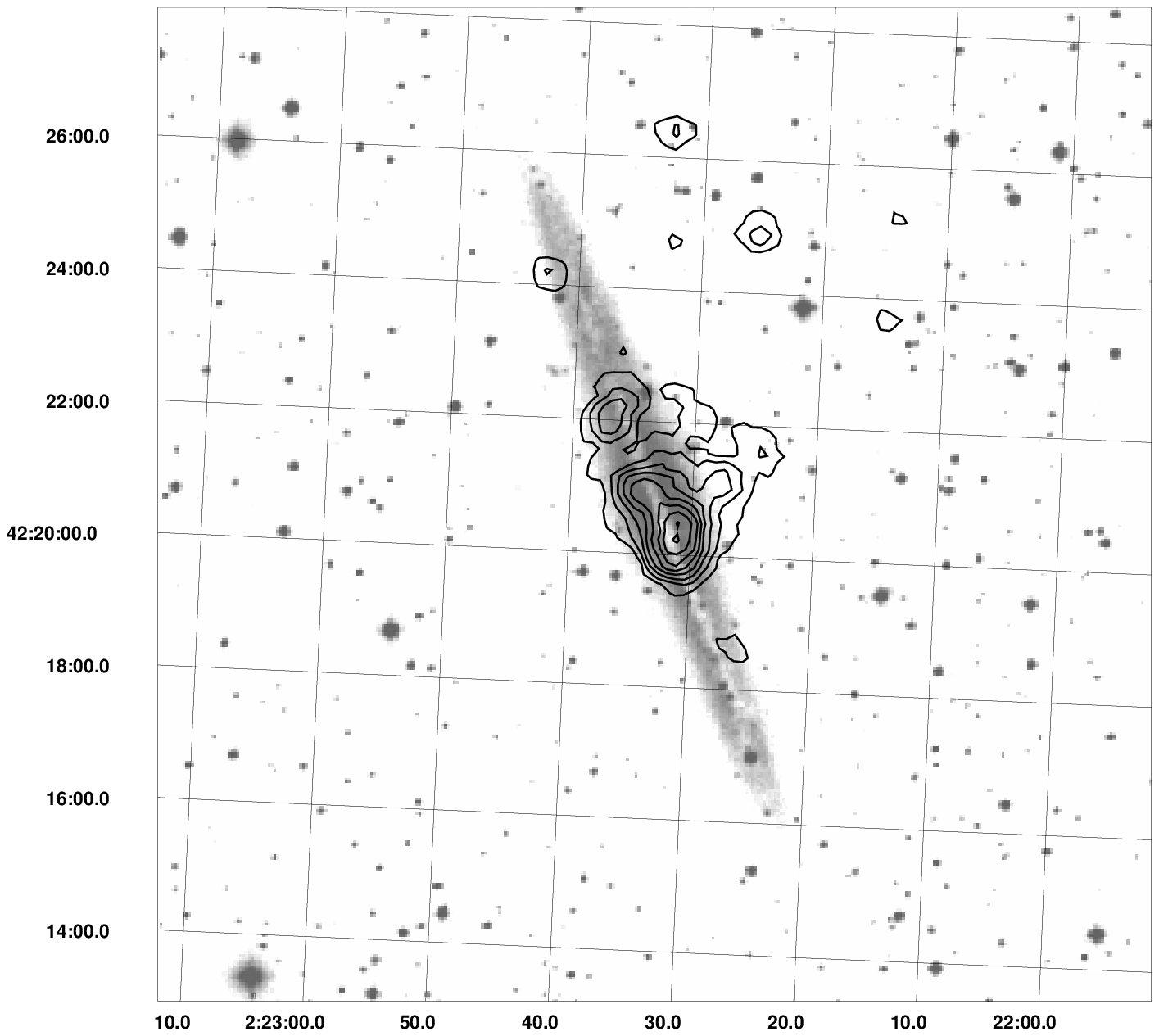,angle=0,width=0.48\hsize,bbllx=70bp,bblly=200bp,bburx=490bp,bbury=580bp,clip=}
\epsfig{file=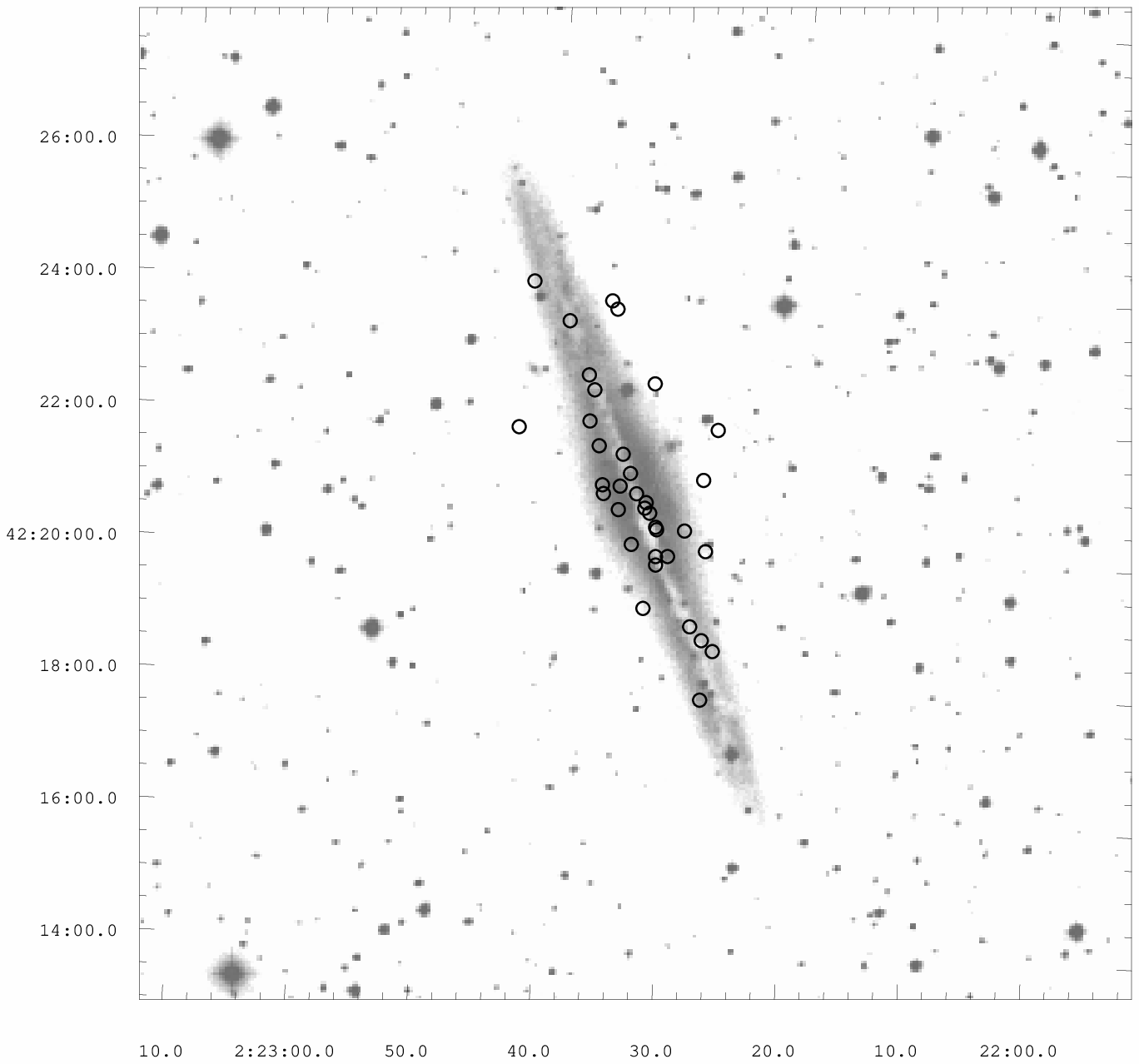,angle=0,width=0.48\hsize,bbllx=70bp,bblly=200bp,bburx=490bp,bbury=580bp,clip=}
\end{flushleft}
\caption{a): The {\sl XMM-Newton} soft X-ray (0.3--2.0~keV) contours
  overlaid onto the optical Digitised Sky Survey (DSS) image. The lowest contour level is set
  at 3$\sigma$ above the background. b): point sources.
\label{fig:diff}}
\end{figure*}

\section{OBSERVATIONS}

NGC\,891 is similar to the Milky Way in optical luminosity
($m_B^{0,i}=9.4$ from RC3, implying $M_B=-20.4$), Hubble type (Sb) and
rotational velocity \citep[225~\kms,][]{b1,b2}.  It is a nearby
edge-on spiral, and thus has been the subject of many detailed studies
of interstellar dust and gas, from observations of the radio continuum
\citep[\eg,][]{dahlem94}, 
H~I \citep[\eg,][]{ssv97}, carbon monoxide
\citep[\eg,][]{sofue93} and molecular hydrogen \citep[\eg,][]{vv99}.

However, there is considerably more star formation in NGC\,891 than in
the Milky Way, presumably due to the presence of about 2.5 times as
much molecular gas \citep{b3}.  As a result,
NGC\,891 is found to be twice as
luminous at IR wavelengths \citep{b4}, and there is evidence of
enhanced dust extinction for optical light, even outside the plane
\citep[\eg,][]{hs97}. The extended gaseous halo of the galaxy in
evident from $H\alpha$ observations using HST/WFPC2, with filaments
reaching up to 2.2~kpc above the galactic plane \citep{rossa04}.

Other wavelengths have yielded other interesting details. Polarised
radio emission has been detected from NGC\,891 \citep{b58},
illustrating the nature of the interstellar magnetic field. A luminous
radio halo has been well established with a scale height of 2.7~kpc
\citep{b59}.

Throughout this analysis, we assume 
a distance of $9.08 \pm 0.45$~Mpc for NGC\,891, which
is a weighted mean of the values obtained by
\citet{b25} and \citet{b26}). Other selected parameters are shown in
Table~\ref{tab:galaxy}. 
Where relevant, we use $H_0$= 72~km~s$^{-1}$~Mpc$^{-1}$.

\begin{table}
  \centering
  \caption{The nearby edge-on spiral galaxy NGC\,891.\label{tab:galaxy}}
  \begin{tabular}{@{}lc}
  \hline
  Parameter  &   Value\\
  \hline
  RA (h m s) & 02:22:33.41 \\
  DEC ($^{\circ}~'~''$) & +40:20:56.9 \\
  Diameter (major) & 14.1$^\prime$ \\
  Diameter (minor) & 3.1$^\prime$ \\
  Major-axis position angle & 22$^\circ$\\
  Redshift & 528 km/s\\
  Adopted Distance & $9.08 \pm 0.45$ Mpc\\
  \hline
\end{tabular}

\begin{flushleft}
Notes: RA and Dec are given in J2000. The diameters correspond the
$D_{25}$ ellipse (i.e. the length of the axis at the isophotal level
of 25 mag/arcsec$^{2}$ in the B Band) Source: {\it NED, LEDAS}. Source
of distance explained in \S2.
\end{flushleft}
\end{table}

\subsection{XMM-Newton observations}

NGC\,891 was observed with the {\sl XMM-Newton} telescope on August 22
2002, as part of a GTO observation. The galaxy was at the centre of
the field of view, with the entire emission also contained within the
field of view. 

The {\sl XMM-Newton} data was observed with three different cameras,
MOS1, MOS2 and PN. Each of the data sets were analysed
separately. After running the standard SAS tasks: {\sc cifbuild} and
{\sc odfingest}; the chains to process the {\em odf}s were run ({\sc
emchain, epchain}) simultaneously running the {\sc badpixfind} command. These
have proved to be considerably better at detecting bad pixels than the
tasks {\sc emproc} and {\sc epproc}. Having extracted the lightcurve
in the 10--15~keV energy range, the observation was found t
contain a
considerable amount of flaring which we used to identify the good time
intervals. The periods of high background were identified by eye at
55/40.0 counts for the MOS/PN cameras respectively. The resulting good
times are 14.6~ks (originally 18.0~ks) for MOS1/MOS2; and 9.6~ks
(originally 15.0~ks) for PN.

Events were filtered based on the the pattern parameter, which
indicates the geometry of the detection of each event, i.e. the number
of adjacent pixels that detect each photon. The events that were
retained in the filtering were the ones that are well-calibrated
(singles, doubles and quadruples for MOS \& singles and doubles for
PN). Due to the extent of the diffuse emission present in this galaxy,
we used blank sky background files for subtraction of our data.  We
followed the technique set out by \cite{b8} for producing images and
spectra that are correctly background subtracted.  

Given the high density of point sources, and the high inclination, it
seemed that manually setting a region with exclusion regions based
around the point sources (extracted from {\sl Chandra} data-- see
below) was more appropriate. The radius of the exclusion region was
set by the SAS task {\sc calview}, using the encircled energy (PSF)
function, at $15''$ (corresponding to a PSF fraction of 75\%). When
the sources were removed, we used the CIAO task {\sc dmfilth} to
interpolate over the holes. The width of the background annuli were
initially taken to be 1.5 times the radius of the source region, but
many were adjusted to ensure that the background annuli did not
overlap with any source regions. In some cases it was impossible to
extract adequate background annuli due to the compact nature of the
sources on each other. In that case, the sources were merged and
treated as a single source removal region.  When analysing the
spectral data, the data from the three cameras were fitted
simultaneously together to ensure the best fit. However, for the
radial profile analysis, the data was mosaicked together using the
task {\sc emosaic}.

\begin{figure*}
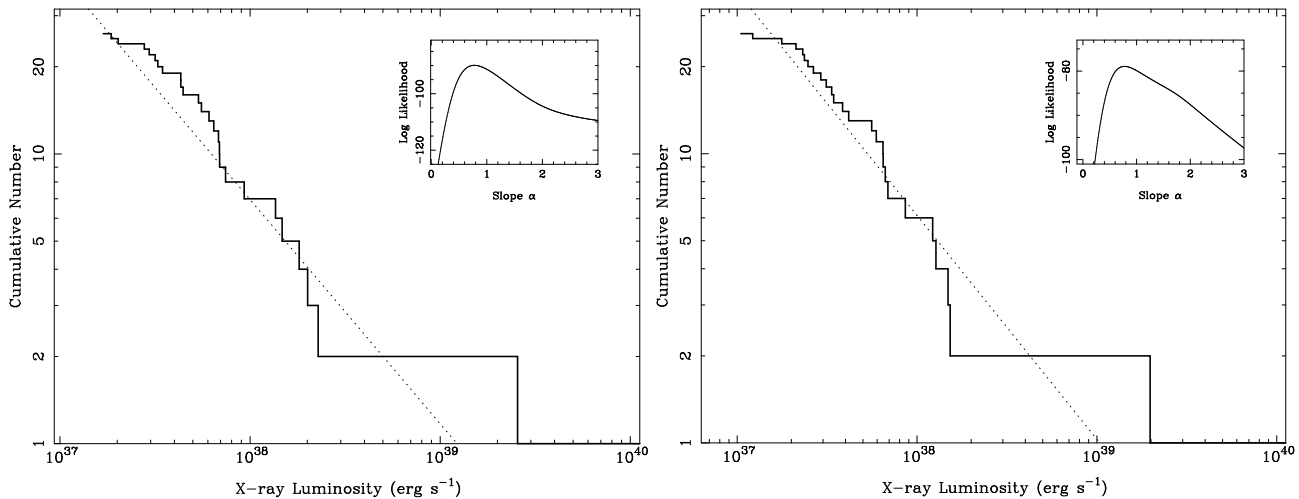

\epsfig{file=trs-fig2a-rev.ps, angle=-90, width=0.48\hsize}
\epsfig{file=trs-fig2b-rev.ps, angle=-90, width=0.48\hsize}
\caption{The X-ray luminosity function (XLF) for point sources,
in NGC~891, in the 0.3--8keV (left) and 2--8 keV (right) energy
ranges, measured from the Chandra observation. The slope, 
defined in (\ref{eq:cumlogn}), is calculated
to be $\alpha=0.77_{-0.10}^{+0.13} $ 
and $\alpha=0.77_{-0.12}^{+0.16} $ respectively. 
Since the slope does not
substantially change for the higher energy range, we conclude that
differential extinction within the inclined spiral does not
significantly affect the shape of the XLF. The inset in each plot 
shows the shape of the log likelihood function,
derived from the method of 
evaluating the slope and errors, discussed in \S3.1.
\label{fig:lumfunct}}
\end{figure*}

\subsection{Chandra observations}

To better characterise the point sources, we used the archived {\sl
Chandra/ACIS-S} observation (51 ks, PI: Bregman), which was reduced in
a similar way, using the CIAO software with online threads. Once the
{\sl Chandra} data preparation was completed, the CIAO tool {\sc
wavdetect} was run on the {\sl Chandra} data.  We narrowed down the
returned list of sources by excluding all the sources outside the
$D_{25}$ ellipse (see Table~1), and considering only the sources
inside. We used the wavelet scales 1, 2 and 4 to extract the point
sources, as these provided the best identification of all the point
sources present. These sources are shown on a DSS overlay in
Fig.~\ref{fig:diff}). We used the results to generate an X-ray flux of
each source to generate a luminosity function
(Fig.~\ref{fig:lumfunct}).

\section{ANALYSIS OF THE X-RAY OBSERVATIONS}

Here we analyse the X-ray observations described in the previous
section to characterise the population of point sources in NGC\,891,
using the archived {\sl Chandra} data set, utilising the superior
resolution of {\sl Chandra}/ACIS setup. We use this observation to
subtract the point sources from the XMM-Newton observation, to
investigate the nature of the diffuse emission.

\subsection{The Point Source Luminosity Function from Chandra}

We detected a total of 26 point sources within the $D_{25}$ ellipse,
down to a flux limit of $10^{-15}$ erg~cm$^{-2}$~s$^{-1}$,
which at the adopted distance of NGC~891 amounts to a luminosity of
$F_{min}=10^{37}$  erg~s$^{-1}$.
The regions were all output as ellipses containing 99.7\% (3$\sigma$)
of the source counts. As a considerable number of the sources
contained few counts, an accurate spectral fit was not possible, so we
modelled each of the point sources individually in XSPEC using an
absorbed power-law model with a slope of $\Gamma=1.8$. The point
sources with more than 100 counts were individually fitted.  Flux
values were established for each of the point sources in an energy
range of 0.3--8.0~keV, and these were combined to make a luminosity
function (Fig.~\ref{fig:lumfunct}). We discuss the point source
associated with SN1986J in greater detail in the following section.

As we will see below, the slope of the luminosity function can be used
as an indicator of recent star formation activity in a galaxy. This
so-called $\log N$-$\log S$ relation is usually expressed in the
cumulative form 
\begin{equation}
\log N(>S) = -\alpha \log S + \kappa
\label{eq:cumlogn}
\end{equation}
where a
straight line is fitted to the cumulative histogram. Here we adopt a
more robust method of measuring the slope of the luminosity function
with the potential of better measuring the uncertainty of the slope.
This is a generalised version of the 
much-used method adapted from \citet{crawford70} and
\citet{murdoch73}, the crucial difference being that in the former,
errors in flux measures are not taken into account and in the latter,
all errors are considered equal. 

Here, each of our $n$ point sources has a measured luminosity $F_i$,
given the adopted distance to the galaxy, and an independently
estimated error
$\sigma_i$.  We represent the probability distribution of the
luminosity $S$ of point sources in the galaxy in its differential form
\begin{equation}
P(S) \,dS = A S^{-\beta}\, dS.
\label{eq:difflogn}
\end{equation}
On comparison with (\ref{eq:cumlogn}), $\alpha=\beta-1$. 
Our exercise thus consists of maximising the log likelihood function
\begin{equation}
\mathcal{L}= \sum^n_{i=1} \ln P(F_i,\sigma_i),
\label{eq:loglike}
\end{equation}
where the distribution
of our 
measured values of flux and standard deviation $(F_i, \sigma_i)$
is given by
\begin{equation}
P(F_i,\sigma_i) = \frac{\int_0^\infty P(F_i,\sigma_i | S)\> P(S) 
        \,dS}{\int_{F_{\rm min}}^\infty 
         \int_0^\infty P(F_i,\sigma_i | S)\> P(S) \,dS\,dF}.
\label{eq:probf}
\end{equation}
Assuming the errors of measuring flux and luminosity are distributed as
a Gaussian, the integrand above is given by
\begin{equation}
P(F_i,\sigma_i | S) P(S)\, dS= \frac{A}{\sigma_i\sqrt{2\pi}}
                 S^{-\beta} e^{-(F_i-S)^2/2\sigma_i^2} \, dS.
\label{eq:prob_integrand}
\end{equation}

We numerically find the value of $\beta$ (thus $\alpha$) for which
$\mathcal{L}$ in (\ref{eq:loglike}) is maximum.  For large $N$, the
probability distribution for $\beta$ is asymptotically Gaussian, 
and the $1\sigma$ error in $\beta$ corresponds to
$\Delta\mathcal{L}=0.5$. For our limiting 0.3-8~keV luminosity,
$F_{min}=10^{37}$  erg~s$^{-1}$, we find the slope of the cumulative
LF given by (\ref{eq:cumlogn})
to be $\alpha=0.77_{-0.10}^{+0.13} $, as
shown in Fig.~\ref{fig:lumfunct}.

Since NGC\,891 is almost exactly an edge-on galaxy, the effect of
differential local extinction might be important within the point
source population, which might lead us to progressively miss the
fainter objects on the far side of the galaxy. Since the effect of
such extinction is energy-dependent (being larger at lower energies,
particularly below 2~keV), we tested whether it is substantial by
re-calculating the luminosity function for the energy range 2--8 keV
(Fig.~\ref{fig:lumfunct}b). The slope was found to be the same at
$\alpha=0.77_{-0.12}^{+0.16} $, leading us to conclude that the effect
of differential extinction is not very serious for this target.

\subsection{The diffuse X-ray emission from XMM-Newton}

We extracted the spectrum of the diffuse emission
in the energy range 0.3--6 keV, excluding 
a circle of 15 arcsec around each detected  point source,
from the 3 {\sl XMM-Newton} camera data and
analysed them with the {\it heasoft} package XSPEC.  We performed
simultaneous fitting of all three data sets in order to fit the best
model. We also tried to add the {\sl Chandra} diffuse emission to the
fits, but due to the limited collecting area of the {\sl Chandra}
telescope, we found the diffuse emission spectra generated did not
improve the fit.

\begin{figure*}
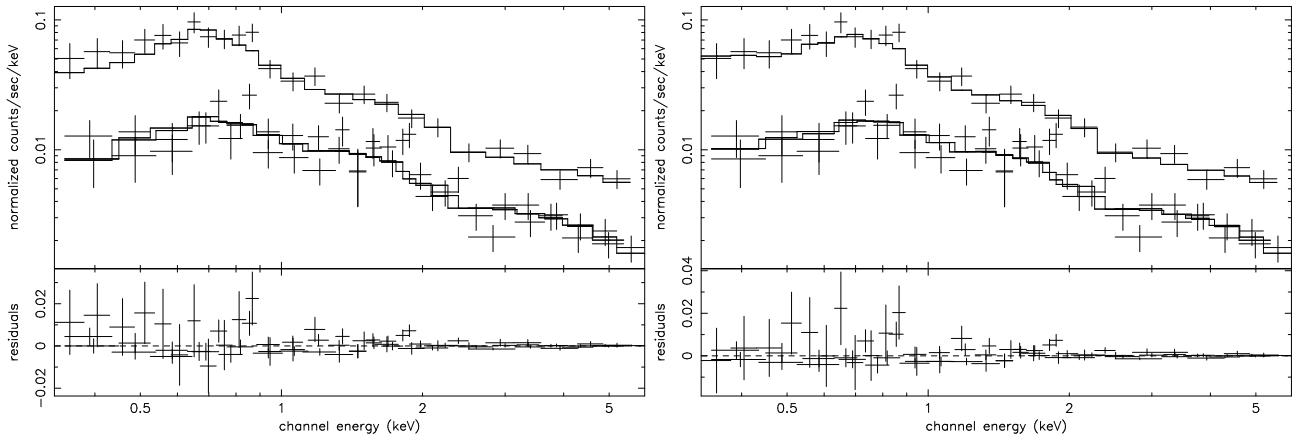

\epsfig{file=trs-fig3a.ps, angle=-90, width=85mm}
\epsfig{file=trs-fig3b.ps, angle=-90, width=85mm}
\caption{(a): The X-ray spectra of NGC\,891, with an absorbed single 
  temperature model + power~law. Spectra for 
  all three {\sl XMM-Newton} cameras have been simultaneously fitted
  in the 0.3--6.0~keV energy range. The top spectrum is from
  the PN camera. (b): As in (a), but the fitted model is an absorbed dual
  temperature + power~law. 
\label{fig:mektot}}
\end{figure*}

We initially fitted an absorbed single temperature plasma model with a
power-law component to the data. We fixed the Galactic value of column
density $n_H$ of
6.78$\times$10$^{20}$ cm$^{-2}$ (obtained from {\sl Colden}).
Since the target is an edge-on galaxy, the local column density is
expected to be substantial- this was looked for, and fixed at a fairly
low value of $n_H$ (local) at 2.0$\times$10$^{19}$ cm$^{-2}$.
The power-law model was added to our data to account for
point source contamination.

Figure~\ref{fig:mektot}a shows the best fit to all three data
sets, having a $\chi^{2}$=68.35 with 65 d.o.f., leading to
$\chi^{2}_{red}$=1.06. We extract a value of $kT=0.26\pm 0.01$~keV.
In comparison, \citet{b6} fit a single temperature model to the 
data with a cooler temperature component ($0.11\pm 0.03$~keV).
\citet{b5} found that the data was best fitted by a two-temperature
model with $kT$ at $0.31$ and $10$~keV. 

\begin{table}
  \centering
  \caption{Summary of recent fits to the diffuse X-ray emission. The 
{\sl XMM-Newton} data are the results generated from this work.
  \label{tab:diff}}
  \begin{tabular}{@{}ccccc}
  \hline
  Obs. & Model & $kT_{1}$ & $kT_{2}$ & $\chi^{2}_{red}$ \\
        &       & (keV)    &  (keV)   &                   \\
  \hline
  {\sl XMM} & pow+mek & $0.26\pm 0.01$ & - & 1.06 \\
  {\sl XMM} & pow+mek+mek & $0.08 \pm 0.01$ & $0.30\pm 0.03$ & 1.05 \\
  {\sl ROSAT}$^{a}$ & mek+mek & 0.31 & 10 & - \\
  {\sl ROSAT}$^{b}$ & mek & $0.11 \pm 0.03$ & - & 1.10 \\
  \hline
\end{tabular}
\begin{flushleft}
Notes: References: a) \citet{b5}, b) \citet{b6}
\end{flushleft}
\end{table}

As the result from \citet{b5} implied a two temperature model, 
a second temperature component was added to the data 
(Fig.~\ref{fig:mektot}b). The two temperatures that were inferred, 
as a result, were  $kT=0.08\pm 
0.01$~keV and $0.30\pm 0.03$~keV. The two temperatures agree within 
errors to the cooler temperature component derived by \citet{b5} and 
the single temperature component derived by \citet{b6}.

However, it should be stated that the spectrum of the data was best 
constrained by a single temperature fit, and the improvement in the 
$\chi^{2}_{red}$ by adding a second temperature component is not 
statistically significant (f-test statistic value $=1.05$).

A summary of our fit parameters compared with 
those of \citet{b5} and \citet{b6} are presented in Table \ref{tab:diff}.

\begin{figure}
\epsfig{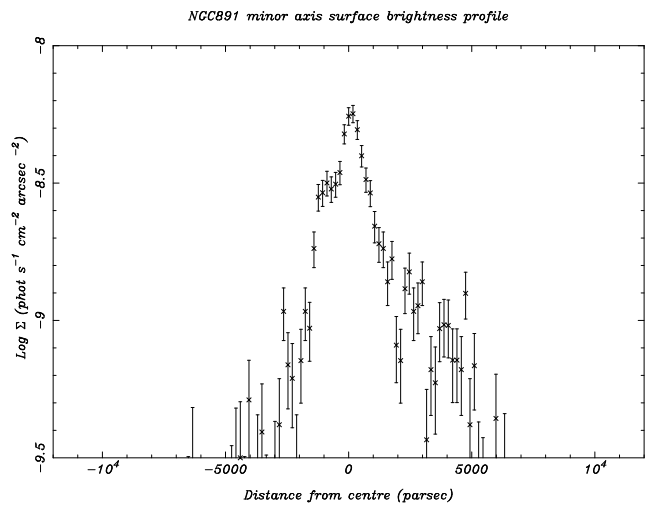}
\caption{Minor axis radial surface brightness
 profile of the diffuse emission. The plot is in the 0.3--8.0~keV
energy band.
\label{fig:radial}}
\end{figure}

Our results do seem to complement the {\sl ROSAT} data to some
extent. Our single and dual temperature model values are consistent with
temperatures that both have derived. \citet{b5} do find a high
temperature component at 10~keV, which is not determined from our
fits. The reason for this may be the poor resolution of {\sl ROSAT},
which means that it is difficult to remove point sources
effectively. This effect is compounded by the high inclination of the
system. Also, SN1986J was considerably brighter, at the time of the
ROSAT observation more than ten years ago, than it is now, and may
well have contaminated the extracted diffuse emission. Finally, the hard 
X-ray flux of a nearby unresolved bright point source near SN1986J 
might also account for this discrepancy \citep{b6}.

We also extracted images of the diffuse emission from the {\sl
XMM-Newton} data. These images complement the X-ray observations from
{\sl ROSAT} \citep{b5} and {\sl Chandra} \citep{b6}. There is a
considerable amount of emission from the centre of the galaxy in the
NW direction. Figure~\ref{fig:diff}a shows this result for the soft
energy band (0.3--2.0~keV). The X-ray contours are overlaid onto the
optical Digitised Sky Survey (DSS) image. The image was smoothed with an
adaptive gaussian program {\sc asmooth} such that the $1\sigma$ width of
the gaussian contained $40/3$ counts around each pixel. We also
took a minor axis diffuse emission profile to show the extent over
which the extended emission protrudes (Fig.~\ref{fig:radial}). The
emission protrudes out to just over 6~kpc from the centre of the
galaxy in the NW direction, but drops off rapidly in the SE direction
outside the plane of the galaxy.

\subsection{Evidence of central AGN}
Using the Chandra observation, \citet{b7} found a weak hard (2.0-8.0~keV) 
X-ray source near the 
nucleus of the galaxy, defined as the position of the radio 
continuum point source \citep{b2}. This was suggested to be a
weak AGN possibly associated with the radio source. Unfortunately,
we were not able to detect this source with our {\sl XMM-Newton} observation,
most likely due to the poorer resolution compared to {\sl Chandra}.

\section{Comparison with other local galaxies}

In order to characterise the global properties of \n891 and find
whether its properties are consistent with that of a starburst galaxy
or not, we compile various X-ray, near-IR and far-IR properties of
several nearby spiral galaxies, of which some are normal galaxies and
some are actively star-forming and starburst galaxies.  Starting from
a sample of galaxies selected from \citet{b15}, a few galaxies from
\citet{b7} and \citet{b15}.  A selected list of parameters are
collected in Table \ref{tab:xlfdata}.

\paragraph*{Far-Infrared luminosities}
We calculated the FIR lumiosities (erg~s$^{-1}$) differently 
depending on whether the source was from {\sl ISO}
(preferably) or {\sl IRAS}. We used the equations from \citet{b14} 
(fluxes in Jy):
\begin{equation}
L_{FIR}=1.89\times 10^{-14} (1.36f_{60}+0.958f_{100}+0.439f_{180}),
\label{eq:FFIR1}
\end{equation}
and for {\sl IRAS}
\begin{equation}
L_{FIR}=1.26\times 10^{-14} (2.58f_{60}+f_{100}),
\label{eq:FFIR2}
\end{equation}

\paragraph*{Near-Infrared luminosities}
To calculate the 
K-band luminosity $L_{K}$ (erg~s$^{-1}$), we derived the following 
expression using the zero point values from \citet{b23} :
\begin{equation}
\log L_{K}=43.13-0.4K_{tot}+2\log D,
\label{eq:LK}
\end{equation}
with $D$ in Mpc, and $K_{tot}$ the total K-band magnitude taken from
the 2MASS galaxy atlas \citep{b60}.  

\paragraph*{X-ray luminosities}
$L_{X}$ was established from the
literature, but different authors used varying energy ranges.
\citet{b7} evaluate their fluxes for an energy range 
0.3--2.0~keV (soft X-ray) We have adopted this energy
range to keep our values for $L_{X}$. In the case where the data was
not in the necessary energy range, we used the Portable Interactive
Multi-Mission Simulator (PIMMS) to generate our data. Using the model
fit parameters for the diffuse emission in the literature, we
estimated a revised flux in the appropriate energy range. 

\paragraph*{Deprojected area of galaxy}
We differ
from the calculation made by \citet{b15} for the area of the $D_{25}$
ellipse. \citet{b15} calculated the galaxy area as $\pi ab\,\cos\theta$. 
For systems that are face on (such as NGC\,4214) this
calculation is adequate.
However, for very high inclination systems, the
measure of the semi-minor axis  simply represents the width of
the disk, and cannot be used in calculating the area. Therefore, for
the sake of consistency we choose to evaluate our area as $\pi a^{2}$
for all galaxies in our sample. 

\subsection{Starburst or not?}

\begin{figure*}
\epsfig{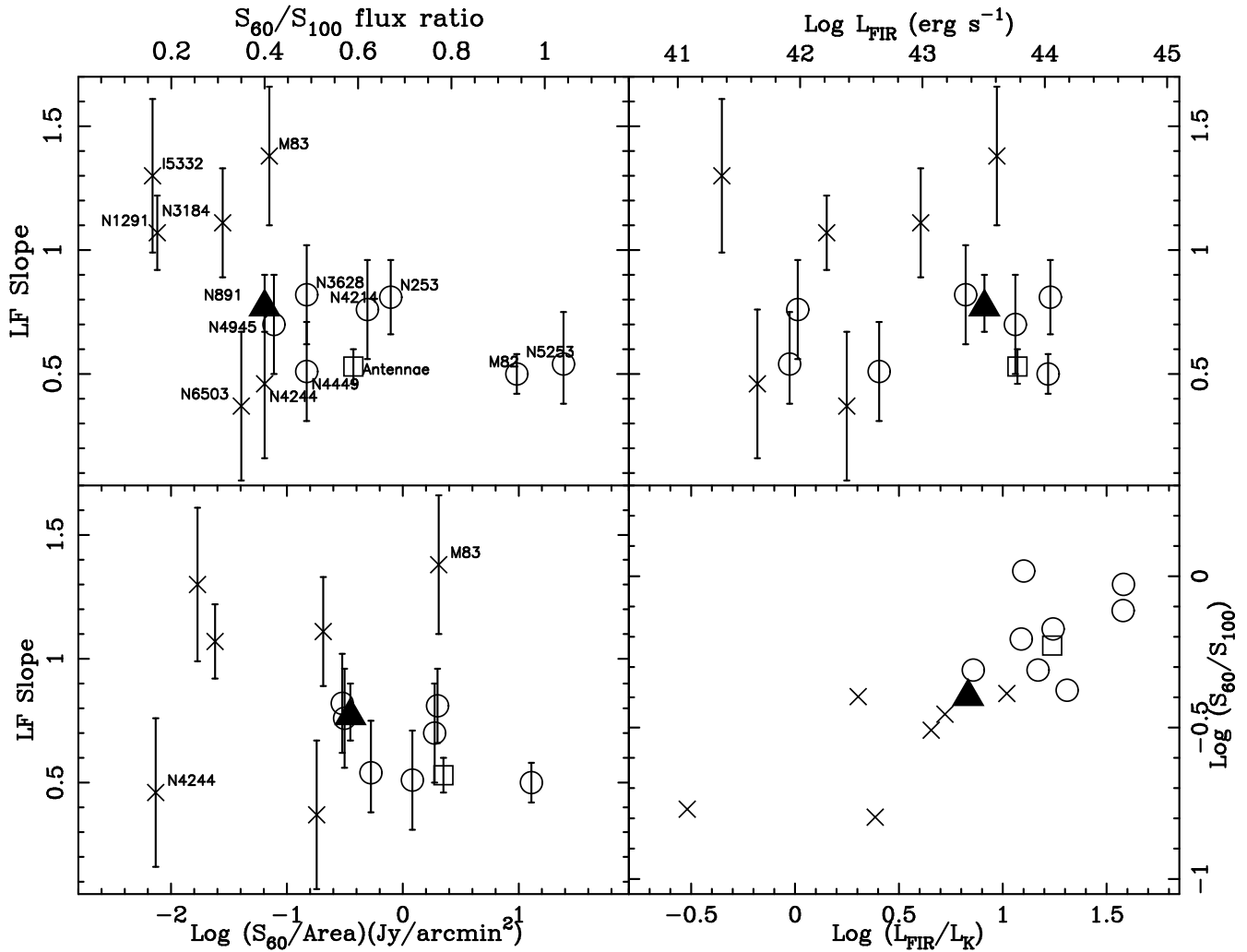}
\caption{Top left a): the XLF slope plotted against the {\sl IRAS} flux ratio
  S$_{60}$/S$_{100}$ for a sample of nearby galaxies. The circles
  represent starburst galaxies; crosses represent normal spirals;
  square for interacting galaxies; and the filled triangle represents
  NGC\,891. The same symbols are used throughout this paper. The data
  labels have been removed for all other plots for neatness. The only
  labels left in are those referred to in the text. The error on NGC\,891
 is the standard deviation, discussed in \S3.1. Top right b): the XLF
  plotted against far infra-red luminosity $L_{FIR}$. Bottom left c):
  the XLF plotted against the ratio of the {\sl IRAS} 60$\mu$ flux and the
  spiral galaxy area. Bottom right d): the {\sl IRAS} flux ratio plotted 
against $L_{FIR}$ normalised with the K-band luminosity. 
\label{fig:xlfplots}}
\end{figure*}

There has been much discussion as to whether NGC\,891 is a normal
spiral galaxy, or whether it is a starburst \citep{b7}. We try to
categorise NGC\,891 by means of comparison with a sample of 16 nearby
galaxies.  Figure~\ref{fig:xlfplots} contains four plots,
which are described below.

The slope of the luminosity function can be used as an
indicator of the general properties of the host galaxy. If the star
formation rate in the galaxy is constant, then the luminosity function
of the sources should be fit by a single unbroken power-law
model. If the galaxy is a starburst one, then new high mass X-ray binaries
(HMXBs) would be formed, breaking the luminosity function slope. The
break in the slope would decrease with time, and would be an
indication of the time of previous bursts in the galaxy. 

As we saw above, in NGC\,891, we chose to fit 
the XLF slope can be fitted by a single power-law.
In the case of NGC\,1482, the galaxy is further away
(22~Mpc) and only two sources could be detected so no XLF slope was
determined. Two other galaxies NGC\,4244 and NGC\,6503 also have low
statistics (with 3 and 4 sources respectively), but XLF slopes were
calculated by \citet{b37}. We include these in our plots, but they do
not appear to be representative of the normal spirals category.

Figure~\ref{fig:xlfplots}a plots the XLF against the $S_{60}/S_{100}$
flux ratio. \citet{b43} analyse the properties of superwinds and
quantitatively discuss the likelihood of generating a superwind with
respect to the infra-red properties of the galaxy. The galaxies with
high luminosity ($L_{IR}\geq 10^{44}$~erg~s$^{-1}$), large infra-red excesses
($L_{IR}/L_{OPT} \geq 2$) and warm far infra-red colours
($S_{60}/S_{100} \geq 0.5$) are most likely to have galactic
superwinds. For the sample selected, the last criterion is satisfied
for all the starburst galaxies. Of the starburst galaxies in our
sample, superwinds are detected in M82 \citep{b45}, NGC\,253
\citep{b44} and NGC\,4945 \citep{b46}. NGC\,5253 \citep{b47} and
NGC\,4449 \citep{b17} exhibit emission from superbubbles. Although
superwinds and superbubbles are both the result of massive star
formation processes within the densest regions of the hot galaxies,
the winds are able to channel the metals produced straight into the
IGM, whereas the superbubbles do not reach the
outskirts of the host galaxies, hence retaining their newly processed
metals, consequently raising the abundance of the ISM \citep{b48}. The
only starburst galaxies not to have superwinds or superbubbles are
NGC\,4214, which is calculated likely to exhibit blowout \citep{b15}
and the Antennae, which is an interacting system, so no superwind has
formed due to disruption from the merging.

Therefore, there are two distinct categories of result: The low XLF
slope galaxies have a higher ratio of warm IR colours than the high
XLF slope galaxies, which are normal spiral galaxies. NGC\,891 appears 
to have the XLF slope of a starburst, but the $S_{60}/S_{100}$
corresponding more to normal spirals than starbursts. 

Figure~\ref{fig:xlfplots}b plots the far infra-red luminosity
($L_{FIR}$) against the XLF slope of the data.  $L_{FIR}$ is an
indicator for how much dust there is in a galaxy, and hence the star
formation rate \citep{b42}. The FIR also contains the most important
cooling lines in the neutral ISM. Therefore, if the XLF slope is
dependent on star formation rate, then, one would expect a correlation
between these two parameters. However, this is not observed in our
sample. Both the starbursts and normal galaxies span a wide
overlapping range of FIR luminosities. The division between spirals
and starbursts is quite clear in this plot, breaking either side of
1. Therefore, an XLF-$L_{FIR}$ dependence cannot be justified on the
basis of these results.

Figure~\ref{fig:xlfplots}c plots the XLF slope against the $S_{60}$
flux scaled against the area of the $D_{25}$ ellipse.The x-axis
therefore represents the star formation rate per unit area. There is a
definite correlation between the data points on this plot. The higher
star formation rate corresponds to starburst galaxies, which we would
expect to see. The major exceptions to this trend are NGC\,4244 and
M83. The XLF slope of NGC\,4244 was determined with only three data
points \citep{b37}, and has an estimated error on the data. M83 should
be classified as a starburst, as observations indicate that there are
massive clusters of stars in the nuclear region \citep{b49}. However,
as discussed in \citet{b13} the star formation rate is low compared to
other starbursts, and the starburst region is confined to a small area
of the galaxy. Therefore, most of the point sources analysed were
taken from a region outside of the starburst region, so the XLF slope
is more representative of the disk population of normal spirals as
opposed to the standard starburst population.

In Fig.~\ref{fig:xlfplots}d, the warm IR colour ratio
S$_{60}$/S$_{100}$ is plotted against the ratio of
$L_{FIR}/L_{K}$. This ratio tells us the extent of the star formation
activity normalised against the stellar types present in the
galaxies. The different galaxy types are clearly grouped separately,
and all the galaxies seem to follow a linear trend. The normal spiral
galaxies have lower star formation rates, which would imply that they
are less likely to generate galactic superwinds. The galaxies with
higher star formation rates are all starburst galaxies with superwinds
(or superbubbles) present. Again, NGC\,891 seems to sit between the
two categories, making it difficult to categorise as either a spiral
or as a starburst.

\subsection{The X-ray Schmidt law}

In a classic paper, \cite{schmidt59} showed that the rate of
Population~I star formation is proportional to the density of
available gas in a galaxy. This has been re-discovered in various
guises over the last few decades. In terms of the X-ray luminosity
of the galaxy, which is related to a density of hot gas, one expects
it to correlate with the star formation rate, which we can characterise by
the total far-IR flux. 

\citet{b51} analyse the effectiveness of the 2--10~keV X-ray
luminosity as a star formation rate indicator, showing
the linear relation between the X-ray luminosity and the far-IR
luminosities, although there are discrepancies between many other
published results. Previous studies using
{\sl ASCA}, {\sl ROSAT/PSPC}
or {\sl EINSTEIN} values would overestimate $L_X$, being unable
to separate out the emission from X-ray binaries and AGN.

\citet{b52} established $L_{X} \propto L_{FIR}^{0.6}$, although
$S_{60}$ made up the x-axis, and $L_{X}$ ranged from
0.5--3.0~keV. Also with the {\sl EINSTEIN} satellite, \citet{b53}
obtained an almost linear fit to their data $L_{X} \propto
L_{FIR}^{0.95 \pm 0.06}$ agreeing quite well with our results within
errors. This result was determined in the 0.5--4.0~keV energy band and
the sample was a selection of starburst and normal
galaxies. \citet{b51} analyses data obtained from the {\sl ASCA}
satellite and for an soft X-ray energy range of 0.5--2.0~keV obtains
$L_{X} \propto L_{FIR}^{0.87 \pm 0.08}$, an identical result to ours
for a very similar energy range (we use 0.3--2.0~keV).
 
\begin{figure}
\epsfig{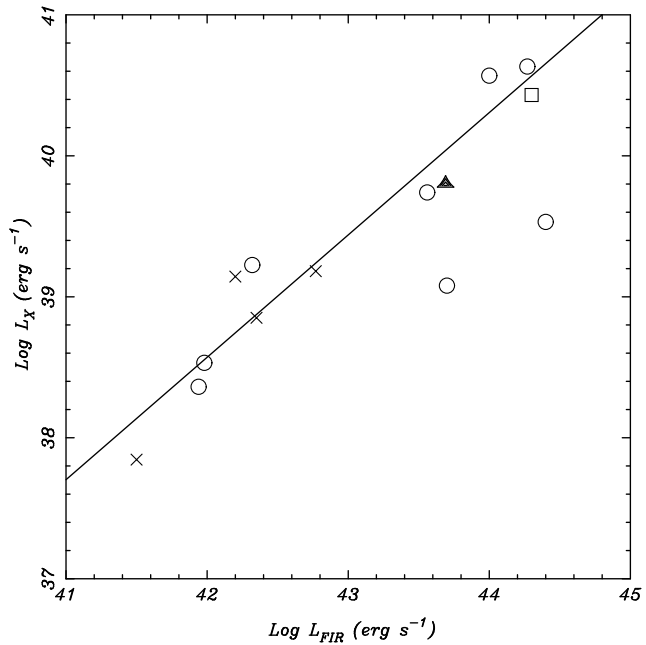}
\caption{The X-ray Luminosity against IR luminosity for a sample of
  galaxies. The circles represent starburst galaxies; crosses
  represent normal spirals; the square for interacting spirals; and
  the filled triangle represents NGC\,891. 
\label{fig:lum5}}
\end{figure}

We plot  $L_{X}$ (the diffuse X-ray luminosity)
against $L_{FIR}$ for our sample galaxies
in Fig.~\ref{fig:lum5}. We establish that the
relation between the data is $L_{X} \propto L_{FIR}^{0.87 \pm
0.07}$, without the two data points on the plot which do not appear to
lie near the line. If these are included,  the fit becomes considerably
worse: $L_{X} \propto L_{FIR}^{0.71 \pm 0.12}$. 
These are the two starburst galaxies NGC\,253 and
NGC\,4945, both of which contain
a superwind emitting from the centre. This would indicate the presence
of an excess of dust in these galaxies containing considerable star
formation. 

\n891 lies comfortably near the best-fitting line, away from
the normal galaxies since its X-ray and far-IR fluxes are higher,
again showing that it does not have as extreme a rate of star formation
as a starburst galaxy.

\section{The resident young X-ray supernova SN1986J}

Even though thousands of supernovae have been found in the optical,
there are instances of only 15 supernovae being detected in the X-ray
\citep{il03,breg03}. Of these, a handful have been monitored over many years,
their X-ray emission dominating their radiative output after about
a year after explosion \citep{pooley02}. Thermal emission from
the reverse shock region 
is expected to be
seen as softer X-ray emission within the expanding SN shell 
as it interacts with the dense stellar wind of the progenitor \citep{cf94}.

One of these Type~II supernovae observed in the X-rays
is SN1986J in \n891,
which was discovered using radio observations from the VLA
\citep{b54}. 
The first X-ray observations of SN1986J were taken with
{\sl ROSAT} \citep{b55}. Since then, two more {\sl ROSAT} and two {\sl
ASCA} observations have been undertaken in the mid 1990's
\citep{b56}. The X-rays are thought to be caused by a shock
propagating through the former stellar envelope \citep{b55}; or due to
a the supernova envelope colliding with a shocked clumpy wind, which
produces the X-rays \citep{b57}. \citet{b55} found that the X-rays
were emitted primarily in the soft X-ray band (0.1--2.5~keV). An
analysis of the {\sl ROSAT} and {\sl ASCA} observations was done by
\citet{b56}, who determined that the X-ray light curve of the data was
dropping by $t^{-2}$, which was considered to be a rather fast decline. 

A summary of all observations are shown in Table \ref{tab:allsn}. Note 
that the X-ray fluxes derived for the {\sl ASCA} data
are an average over all their different models, and so the error shown
here is simply the standard deviation of these data sets.

\begin{figure}
\epsfig{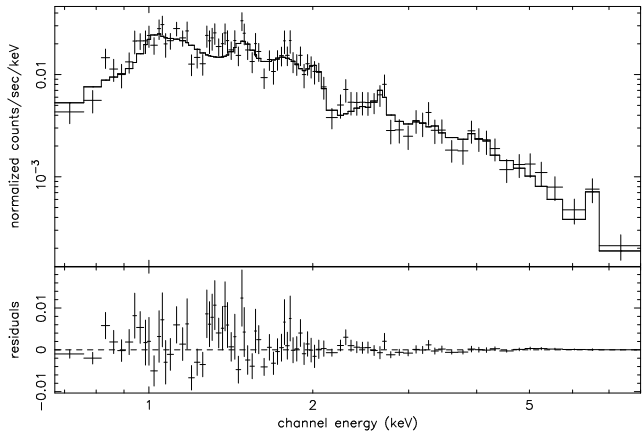}
\caption{The {\sl Chandra} X-ray spectrum for SN1986J.
\label{fig:ch_sn}}
\end{figure}

\subsection{The spectrum of SN1986J}
The spectrum for the {\sl Chandra} observation was extracted using a 
local background and we fitted
different temperature models to the data. Firstly, an absorbed single
temperature thermal plasma model ({\sc mekal}) was fitted with a local and 
Galactic
absorbing column \citep{b55}. The Galactic value for $\log n_{H}$ 
was fixed at $20.8$ ($n_H$ in cm$^{-2}$) and the local absorbing column was
allowed to vary. The abundance was initially fixed at 0.3 then 1.0 
solar,
and finally was allowed to vary. The local $\log n_{H}$ was evaluated
at 21.6, in good agreement with the value of \citet{b55}
(21.7), but the temperature range that we obtained from the
model was 3.5-4.1~keV. This temperature component is much larger than
the 2.0~keV values obtained from the ROSAT observations, but smaller 
than the ASCA temperature ranges. A {\sc vmekal} model was also tried, where
the individual element abundances are allowed to vary, as the mekal 
fit results were leaving several emission lines unidentified. The 
result from the vmekal fit is plotted in Fig. \ref{fig:ch_sn}. The
temperature was determined to be 4.6~keV, local $\log n_{H}$=21.5,
and X-ray flux 0.45$^{+0.1}_{-0.7}\times 10^{-14}$ erg s$^{-1}$cm$^{-2}$. 
Many elements were identified in the spectrum of the supernova, most 
notably, the Fe emission line at 6.4~keV. The abundances are tabulated 
in Table \ref{tab:vmek}. 

On inspection of Fig. \ref{fig:ch_sn},  some of the emission lines are
not fitted that well by a {\sc vmekal} model. Several methods were tried 
to improve the fit, including a two temperature model 
({\sc mekal+vmekal}) and a different single temperature thermal 
plasma model ({\sc apec}). In each of these models, the 
hard temperature and local $n_{H}$ were similar to the results obtained from 
the single temperature fitting. For the two temperature fitting, the lower
 temperature component was 0.11~keV which is likely the component from the 
diffuse emission of the galaxy, and not the supernova. The results from all 
the Chandra fits are tabulated in Table \ref{tab:snfits}. None of the fits 
identified any of the other emission lines. As all the flux values generated 
from the fits were similar, the value adopted for our supernova flux is the 
weighted mean of the different models: 
9.6$^{+0.6}_{-0.4}\times 10^{14}$~erg~s$^{-1}$cm$^{-2}$. The energy range 
chosen was 0.5-2.5~keV to be consistent
with the energy ranges used in the ASCA and ROSAT data. 

For the {\sl XMM-Newton} data, the flux was much harder to
determine. As the flux is dropping every year, the supernova becomes
harder to model the later the observations are. We fitted all three
cameras simultaneously to try and obtain the best fit to the data. Our
data was best fitted with a single temperature model ({\sc mekal}) with a
temperature of 3.6~keV and absorbing column density $\log
n_{H}=21.53$. This fit resulted in a flux of $8.31\pm 1.06 \times
10^{14}$~erg~s$^{-1}$~cm$^{-2}$. Fitting a dual temperature model did not 
improve the fit. The temperatures obtained from the fit were 0.21~keV and 
3.82~keV for the soft and hard components respectively. This fit did produce 
a surprisingly large local $n_{H}$ of 22.6. The two-temperature
flux was 8.40$\pm 1.16 \times 10^{-14}$~erg~s$^{-1}$cm$^{-2}$. The value
adopted from the flux 
is the mean of the four different temperature fits for the {\sl XMM-Newton} 
data: 8.5$\pm 0.15 \times 10^{-14}$~erg~s$^{-1}$cm$^{-2}$.

\begin{table}
 \centering
  \caption{Summary of X-ray observations of SN1986J. \label{tab:allsn}}
  \begin{tabular}{@{}cccc@{}}
  \hline
   Telescope & Date & MJD & Flux \\
   (1) & (2) & (3) & (4) \\
\hline
{\sl ROSAT}   & Aug 1991 & 48486.1 & $7.90 \pm 0.41$ \\
{\sl ROSAT}   & Jul 1993 & 49206.5 & $4.91 \pm 0.24$ \\
{\sl ASCA}    & Jan 1994 & 49374.5 & $6.07 \pm 0.13$ \\
{\sl ROSAT}   & Jan 1995 & 49744.0 & $3.95 \pm 0.26$ \\
{\sl ASCA}    & Jan 1996 & 50113.5 & $5.28 \pm 0.10$ \\
{\sl Chandra} & Nov 2000 & 51852.0 & $0.96^{+0.05}_{-0.06}$ \\
{\sl XMM-Newton}     & Aug 2002 & 52508.0 & $0.85 \pm 0.15$ \\

\hline \\
\end{tabular}
\\
\begin{flushleft}
NOTES: Column (1): The first 2 {\sl ROSAT} observations were taken
with the PSPC, the last with the HRI. Column (4): The X-ray fluxes are
measured in $10^{-13}$~erg~s$^{-1}$~cm$^{-2}$. The {\sl ASCA} and {\sl
  ROSAT} data are compiled from \citet{b56}; the {\sl Chandra} and
{\sl XMM-Newton} fluxes are from this paper.
\end{flushleft}
\end{table}

\begin{table}
  \centering
  \caption{Metal abundances for single {\sc vmekal} model.\label{tab:vmek}}
  \begin{tabular}{@{}cc}
  \hline
  Metal & Abundance \\
  \hline
  Ne & 5.12\\
  Mg & 2.71\\
  Al & 5.42\\
  Si & 1.87\\
  S  & 3.8\\
  Ar & 1.09\\
  Ca & 3.48\\
  Fe & 0.92\\

  \hline
\end{tabular}
\\
NOTES:  Abundances are relative to solar.
\end{table}

\subsection{The evolution of SN1986J with time}

\begin{figure}
\epsfig{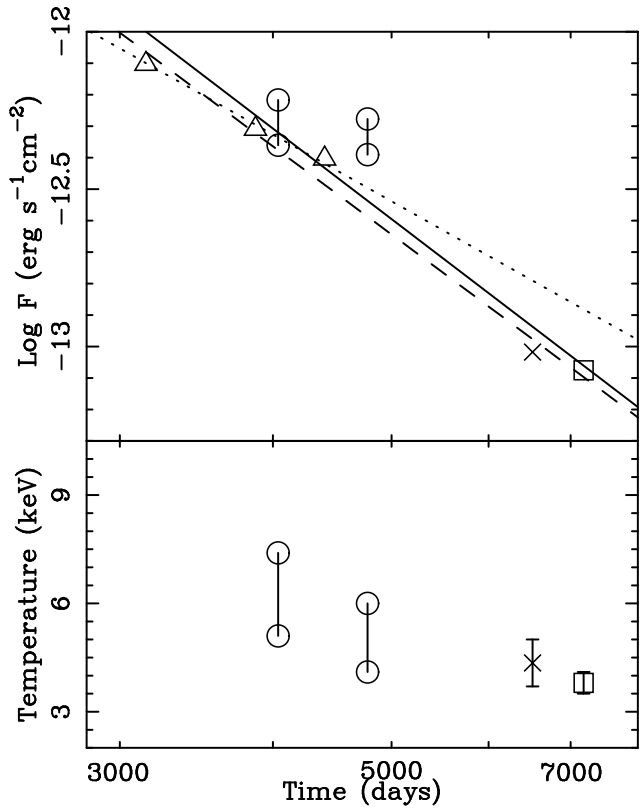}
\caption{The X-ray lightcurve of SN1986J. {\sl ROSAT} data are shown by
  triangles; {\sl ASCA} data by circles; {\sl Chandra} data denoted by
  a cross; and {\sl XMM-Newton} by a square. The {\sl ASCA} data has
  been adjusted to remove point source contributions (see text). The 
lower values represent the point source corrected results.  The
  x-axis represents the time in days since the supernova went off
  ($t=0$, which is taken to be 1983.0). All data is in the soft energy
  range 0.5--2.5~keV. The solid line is the best fit line for all the
  data with adjusted {\sl ASCA} data points. The dashed line is the
  best fit for the data excluding all {\sl ASCA} points. The dotted
line is the best fit from \citet{b56} for ROSAT only. 
Bottom panel: Temperature.
\label{fig:sn_lc}}
\end{figure}

SN1986J has been observed with ROSAT, ASCA, Chandra and XMM over
the last twelve years.
In Fig.~\ref{fig:sn_lc} we plot the 0.5--2.5~keV flux
against time over this period. We recall that from the ROSAT
and ASCA data \citep{b56}, the decline of this flux was found
to be $\propto t^{-2}$, which was considered to be rather steep.

Here we attempt to find a mean rate of decline over a much larger time
scale.  The data from {\sl ASCA} and {\sl ROSAT} appear to be
inconsistent \citep{b56} and so we have established a best fit line 
inclusive and
exclusive of the {\sl ASCA} data. The slopes are calculated as
$-2.99\pm 0.45$ and $-2.89\pm 0.19$ respectively.  In
Fig.~\ref{fig:sn_lc}, we also show an alternative set of points for
ASCA where we attempt to correct for the poor resolution by
subtracting from the the flux of the SN the combined flux of all the
other point sources that are expected to be within its point-spread
function.

However, our results imply $L_{X} \propto t^{-3}$. We can compare this
with the expected behaviour from the work of \cite{cf94}, who show
that $L_X\propto t^{-1}$ when free-free emission dominates ($T>4\times
10^7$K), and $L_X\propto t^{-0.7}$ when line emission dominates
($10^5<T<4\times 10^7$K). Such a trend in the decline in luminosity
is seen in various other Type~II SNe, including SN 1993J in M81
and SN1999em in NGC~1637 \citep{il03}. However,
our measured temperature of the plasma is
around 4~keV, which corresponds to the former range, which indicates
that the decline of $L_X$ is unusually steep in this case.  It is
likely that the poor resolution of the {\sl ASCA} and {\sl ROSAT}
satellites might not have been able to fully eliminate point source
contamination, thus overestimating the fluxes of the data.

\begin{table}
  \centering
  \caption{XMM and Chandra single-temperature 
fits to the supernova lightcurve.\label{tab:snfits}}
  \begin{tabular}{@{}cccccccc}
  \hline
  Telescope & Model & kT & nH & Flux & $\chi^{2}_{red}$ & Abund\\
  (1) & (2) & (3) & (4) & (5) & (6) & (7) \\
	      \hline

   Chandra  & mek        & 3.5 & 0.41 & 9.7 & 1.34 & 0.3F \\
   Chandra  & mek        & 4.0 & 0.39 & 9.6 & 1.16  & 1.0F \\
   Chandra  & mek        & 4.1 & 0.39 & 9.7 & 1.15  & 1.30 \\
   Chandra  & vmek       & 4.6 & 0.30 & 9.5 & 1.09  & -    \\
   Chandra  & mek+vmek   & 4.6 & 0.34 & 9.5 & 1.13  & -    \\
   Chandra  & apec         & 4.0 & 0.39 & 9.9 & 1.39 & 1.48 \\
   Chandra  & vapec        & 4.1 & 0.37 & 9.6 & 1.18  & -    \\
   Chandra  & apec+vapec   & 5.9 & 0.45 & 9.5 & 1.16  & -    \\
   XMM      & mek        & 3.6  & 0.35 & 8.3 & 0.92 & 0.77 \\
   XMM      & apec         & 3.4  & 0.37 & 8.6 & 0.92 & 0.77 \\
   XMM      & vmek       & 3.8  & 1.15 & 8.4 & 0.90 & - \\
   XMM      & mek+mek      & 4.2  & 0.38 & 8.7 & 0.84 & - \\
  \hline
\end{tabular}
\\
\begin{flushleft}
NOTES: Col (3): Temperature component of the supernova
fitting in keV. For two-temperature models this is the harder component.
Col (4): Absorbing column parameter local to NGC891
$\times 10^{22}$cm$^{-2}$ 
Col (5): X-ray flux $\times 10^{-14}$erg~s$^{-1}$cm$^{-2}$. Col (7):
Abundance (relative to solar) from different fitting models. 'F' indicates
a fixed fitting value. These are not shown for the vmekal/vapec model.
\end{flushleft}
\end{table}

\section{Conclusions}

In this paper, we have presented a GTO {\it XMM-Newton} observation of
NGC\,891, a nearby edge-on spiral galaxy. We find that the diffuse X-ray
emission protrudes from the disk in the NW direction out to approximately
6~kpc.  The extraplanar hot gas was previously observed in ROSAT
observations, but the XMM observation seems to show a sharp
cut-off to this gas, showing that this might be the extent to which the
gas has reached. We also find that the best fitting thermal plasma model
seems to require two temperatures of 0.08 and 0.3~keV respectively, though
the fit of a single-temperature plasma of 0.26~keV isn't much worse.
This contradicts previous findings based on ROSAT observations
\citep{b5,b8}.

We use an archived {\sl Chandra} observation to study the point source
population within the $D_25$ ellipse of the galaxy. We use a robust
maximum likelihood method to determine the slope of the cumulative
luminosity function $N(>S) = S^{-\alpha}$, and find that the slope is
rather shallow, $\alpha= 0.77^{+0.13}_{-0.10}$. We have verified that this
isn't predominantly due to extinction, by plotting the XLF for the energy
range $>2$~keV, without much change on slope.

Using a sample of other local galaxies, we have compared the X-ray and
infrared properties of NGC\,891 with those of nearby 'normal' and
starburst spiral galaxies. We conclude that NGC\,891 has more abundant
star formation than a normal spiral, but does not have as extreme
properties as starburst galaxies like NGC\,253 and NGC\,4945, and
that it is most likely a starburst galaxy in a quiescent state.

We examine an X-ray version of the ``Schmidt Law'', which correlates
the rate of star formation in a galaxy to the mass or density of its
available gas. We show that the diffuse X-ray
luminosity, an indicator of gas mass,
of nearby spirals scales with their far infra-red luminosity,
as indicator of dust mass and star formation rate, as
$L_{X}\propto L_{FIR} ^{0.87 \pm 0.07}$, except for extreme
starbursts, and NGC\,891 does not fall in the latter category.

We study the supernova SN1986J in NGC~891 in both {\sl XMM-Newton} and
{\sl Chandra} observations, nearly twenty years after its explosion. It
was studied in the X-ray using ROSAT and ASCA almost a decade ago.
The flux and temperature calculated in those studies of low
spatial resolution could have
significant contamination from their point sources and the diffuse
emission. A direct comparison shows that the temperature of the SN remnant
at the time of observation was $5\times 10^7$~K, which is less than it was
a decade ago, but it also reveals that the X-ray luminosity has been
declining with time ($L_X\propto t^{-3}$) far more steeply than expected.

\begin{table*}
 \centering
 \begin{minipage}{160mm}
  \caption{Luminosity function and flux data for a sample of nearby
  galaxies.\label{tab:xlfdata}} 
  \begin{tabular}{@{}lrlrlrlrccccccl@{}}
  \hline

   Galaxy & \multicolumn{2}{c}{Distance} & \multicolumn{2}{c}{LF slope} & \multicolumn{2}{c}{100$\,\umu$m flux} &
   Source & $S_{60}/S_{100}$ & $\log L_{FIR}$ & $\log L_{B}$ & \multicolumn{2}{c}{$\log L_{X}$} & Area\\
        & \multicolumn{2}{c}{(Mpc)} & & & \multicolumn{2}{c}{(Jy)} & &  & (erg s$^{-1}$) & (erg s$^{-1}$)
	& \multicolumn{2}{c}{(erg s$^{-1}$)} & (arcmin$^{2}$) \\
    (1) & \multicolumn{2}{c}{(2)} & \multicolumn{2}{c}{(3)} & \multicolumn{2}{c}{(4)} & &(5) & (6) & (7) &
    \multicolumn{2}{c}{(8)} & (9)\\
 \hline
N253  &  3.94   & n & $0.81\pm 0.15$         & e & 1770.0 & l & {\sl ISO}  & 0.67 & 43.70 & 42.80 & 39.08 & y  & 595.68 \\
N891  &  9.08   &   & $0.77^{+0.13}_{-0.10}$         &   & 126.0  & m & {\sl ISO}  & 0.40 & 43.69 & 42.67 & 39.78 & z  & 143.14 \\
N1291 &  8.6    & q & $1.07\pm 0.15$         & e & 10.1   & d & {\sl IRAS} & 0.17 & 42.20 & 42.74 & 39.14 & f  & 74.47  \\
N1482 &  22.0   & a & -                      &   & 45.8   & a & {\sl ISO}  & 0.77 & 44.0  & 42.42 & 40.57 & z  & 5.06   \\
N3184 &  11.12  & r & $1.11\pm 0.22$         & e & 29.0   & c & {\sl IRAS} & 0.31 & 42.77 & 42.33 & 39.18 & ab & 42.24  \\
N3628 &  7.91   & s & $0.82\pm 0.20:$        & b & 106.0  & c & {\sl IRAS} & 0.49 & 43.56 & 42.50 & 39.74 & z  & 172.03 \\
N4214 &  2.94   & o & $0.76\pm 0.20$         & g & 29.0   & c & {\sl IRAS} & 0.62 & 41.98 & 40.89 & 38.53 & g  & 59.52  \\
N4244 &  4.49   & p & $0.46\pm 0.30:$        & b & 208.7  & c & {\sl ISO}  & 0.40 & 41.65 & 41.35 & 37.85 & z  & 216.95 \\
N4449 &  4.21   & p & $0.51\pm 0.20:$        & i & 73.0   & h & {\sl IRAS} & 0.49 & 43.05 & 41.48 & 39.23 & i  & 37.82  \\
N4945 &  3.60   & p & $0.70\pm 0.20:$        & b & 1415.5 & d & {\sl IRAS} & 0.42 & 44.40 & 42.45 & 39.53 & z  & 93.65  \\
N5253 &  3.14   & t & $0.54^{+0.21}_{-0.16}$ & k & 29.4   & j & {\sl IRAS} & 1.04 & 41.94 & 40.81 & 38.35 & k  & 19.79  \\
N6503 &  5.27   & p & $0.37\pm 0.30:$        & b & 28.9   & c & {\sl ISO}  & 0.35 & 42.38 & 41.66 & 38.85 & z  & 20.47  \\
M82   &  3.89   & u & $0.50\pm 0.08$         & e & 1351.1 & d & {\sl IRAS} & 0.94 & 44.27 & 42.45 & 40.63 & z  & 98.87  \\
M83   &  4.51   & x & $1.38\pm 0.28$         & e & 638.6  & d & {\sl IRAS} & 0.41 & 43.65 & 42.59 & -     &    & 127.32 \\
IC5332 & 4.63   & v & $1.30\pm 0.31$         & e & 5.1    & d & {\sl IRAS} & 0.16 & 41.87 & 40.97 & -     &    & 39.98  \\
Antennae & 13.8 & w & $0.53\pm 0.07$         & e & 82.0   & c & {\sl IRAS} & 0.59 & 44.30 & 42.54 & 40.43 & aa & 22.05  \\
 \hline
\end{tabular}
NOTES.- Column~(2): distance in Mpc, see text in \S2 for distance
discussion for NGC\,891. Column~(3): modulus of slope generated from
the luminosity function. Data with a : has had the error
estimated. Column~(4): {\sl ISO/IRAS} $100\umu$m flux in
Jansky. Column~(5): Ratio of {\sl IRAS/ISO} 60 to 100 $\umu$m flux
ratio. Column (6): Far IR luminosity calculated using Eqs~\ref{eq:FFIR1} 
and \ref{eq:FFIR2}.  Column~(7): K-band luminosity
calculated using equation \ref{eq:LK}. Column~(8): the X-ray
luminosity ($\log L_{X}$) estimated in the 0.3--2.0~keV
energy range. 
Column~(9): Deprojected Galaxy area = $\pi a^{2}$, where $a$ is the
semi-major axis of the $D_{25}$ isophotal ellipse. \\ 
REFERENCES.-a) \citet{b9}, b) \citet{b37}, c) \citet{b11}, d)
\citet{b12}, e) \citet{b13}, f) \citet{b38}, g) \citet{b15}, h)
\citet{b16}, i) \citet{b17}, j) \citet{b18}, k) \citet{b19}, l)
\citet{b20}, m) \citet{b21}, n) \citet{b24}, o) \citet{b30}, p)
\citet{b31}, q) \citet{b27}, r) \citet{b28}, s) \citet{b29}, t)
\citet{b32}, u) \citet{b33}, v) \citet{b34}, w) \citet{b35}, x)
\citet{b36}, y) \citet{b39}, z) \citet{b7}, aa) \citet{b40}, ab) \citet{b41}.
\end{minipage}
\end{table*}

\section*{Acknowledgements}
Our thanks to Trevor Ponman and Andrew Read for sharing their XMM GTO data
with us, and their advice.  We thank Ben Maughan, Irini Sakelliou and Ed
Colbert for useful comments and help in data analysis, and Joel Bregman for
a very interesting discussion. We also thank an anonymous referee for
useful comments. This publication makes use of data
products from the Two Micron All Sky Survey, which is a joint project
of the University of Massachusetts and the Infrared Processing and
Analysis Center/California Institute of Technology, funded by the
National Aeronautics and Space Administration and the National Science
Foundation.


\label{lastpage}

\end{document}